\documentclass[twocolumn,showpacs,preprintnumbers,amsmath,%
amssymb,nofootinbib,floatfix,prd]{revtex4-1}
\usepackage{graphicx}
\usepackage{calrsfs}
\usepackage[hypertex]{hyperref}

\begin{document}
%\selectlanguage{english}

\def\b{\boldsymbol}
\def\p{\partial}
\def\vt{\vartheta}
\def\e{\epsilon}
\def\ts{\textstyle}
\def\k{\varkappa}
\def\d{\delta}
\def\be{\begin{equation}}
\def\ee{\end{equation}}

\def\apss{Ap\&SS}
\def\mnras{MNRAS}
\def\jgr{J. Geophys. Res.}

\title{Small pitch-angle magnetobremsstrahlung in inhomogeneous curved magnetic fields}

\author{S.R.~Kelner}
\altaffiliation{National Research Nuclear University (MEPHI), Kashirskoe shosse
31, 115409 Moscow, Russia; Max-Planck-Institut f\"ur Kernphysik,
Saupfercheckweg 1, D-6917 Heidelberg, Germany} \email{skelner@rambler.ru}

\author{F.A.~Aharonian}
\altaffiliation{Dublin Institute for Advanced Studies, 31 Fitzwilliam Place, Dublin 2, Ireland;
Max-Planck-Institut f\"ur Kernphysik,
Saupfercheckweg 1, D-6917 Heidelberg, Germany}
\email{Felix.Aharonian@mpi-hd.mpg.de}

\begin{abstract}
The character of radiation of relativistic charged particles in strong
magnetic fields largely depends on the disposition of
particle trajectories relative to the field lines. The motion of particles with trajectories
close to the curved magnetic lines is usually referred to the so-called curvature radiation.
The latter is treated within the formalism of synchrotron radiation
by replacing the particle Larmor radius with the curvature radius of the field lines.
However, even at small pitch angles, the curvatures of the particle trajectory
and the field line may differ significantly.
Moreover, as we show in this paper the trajectory curvature
varies with time, i.e. the process has a stochastic character. Therefore for calculations of
observable characteristics of radiation by an ensemble of particles, the radiation intensities
should be averaged over time. In this paper, for determination of particle trajectories we use
the Hamiltonian formalism, and show that close to curved magnetic lines, for
the given configuration of the magnetic field, the initial point and particle energy,
always exist a {\it smooth trajectory} without fast oscillations of the curvature radius.
This is the trajectory which is responsible for the curvature radiation.
The realization of this regime requires the initial particle velocity to be
directed strictly along the smooth trajectory. If initially the velocity vector
is adjacent to the field line or close to it,
a transition of the particle to the smooth trajectory on a later stage cannot be excluded,
however under certain conditions this can happen only after the particle has already
radiated out all its energy.
Even for an initial narrow angular
distribution of the particle beam,
the energy spectrum of the resulting radiation appears
significantly harder compared to the energy spectrum of the curvature radiation.
This result might have direct relation to the recent spectral measurements of gamma-radiation
of pulsars by the {\it Fermi} Gamma-ray Space Telescope.

\end{abstract}


\maketitle

\section{Introduction}
The motion of a relativistic charged particle in a strong magnetic field is accompanied
by a radiation which in general terms can be called magnetobremsstrahlung.
While moving in a homogeneous field, the particle energy is
gradually radiated away, however, its velocity component $v_{\parallel}^{}$
parallel to the magnetic field remains unchanged. Indeed, this statement is obvious for
$v_{\parallel}^{}=0$, therefore in an arbitrary coordinate system which moves along (or opposite)
the magnetic field, we have $v_{\parallel}^{}={\rm const}$. Therefore,
\be\label{in1}
\frac{c^2}{v_{\parallel}^2}-1=\frac{p_\perp^2+m^2c^2}{p_{\parallel}^2} = {\rm const}\,,
\ee
where $p_{\parallel}^{}$ and $p_{\perp}^{}$ are the parallel and perpendicular
components of the momentum, respectively. This equation allows us
to find the remaining energy of the particle, after the radiative damping
of the perpendicular component of motion. Let us denote
the initial and final values of the parallel and perpendicular
components of the momentum as $p_{\parallel}^{} $,
$p_{\perp}^{}$, and $p'_{\parallel}$,
$p'_{\perp}=0$, respectively.
From Eq.(\ref{in1}) follows that
\be\label{in2}
p'_{\parallel}=p_{\parallel}^{}\,\frac{mc}{\sqrt{p_{\perp}^2+m^2c^2}}\,.
\ee
If the initial perpendicular momentum is nonrelativistic, $p_{\perp}^{}\ll mc$, then
$p'_{\parallel}\approx p_{\parallel}^{}$, i.e. the damping of the perpendicular motion
does not have an impact on the particle energy
However, for $mc\ll p_{\perp}^{}\ll p_{\parallel}^{}$, we have $p'_{\parallel}\ll p_{\parallel}^{}$, i.e.
during the damping of the perpendicular motion the particle loses almost all its energy
At $p_{\perp}^{}\gg mc$, the momenta $p_{\parallel}^{}$ and
$p_{\perp}^{}$ vary similarly with time since the ratio
$p_{\perp}^{}/p_{\parallel}^{}={\rm const}$ (see Eq.(\ref{in2}).
The constant ratio of $p_{\perp}^{}/p_{\parallel}^{}$ has the following explanation.
The ultrarelativistic particle emits photons within the angle $\lesssim1/\gamma$ in the direction of its momentum $\b p$, therefore in the case of $1/\gamma\ll
p_{\perp}^{}/p_{\parallel}^{}$, one may assume that the recoil momentum
is directed opposite to the vector $\b p$.

The radiation of charged particles in strong magnetic fields is determined by
the character of their motion, in particular by the disposition of
particle trajectories relative to the field lines. As we show in this paper, in the close vicinity of
curved field lines should exist a special trajectory the particle motion along which
is not accompanied by perpendicular oscillation. We call this trajectory as {\it smooth trajectory}.
In the case of homogeneous magnetic field the smooth trajectory is a straight line parallel to the
field line. The initial short interval of the smooth trajectory in the curved magnetic field
can be treated as a circle with a radius close to
the curvature radius of the field line. An ultrarelativistic particle moves steadily
along this circle, therefore for calculations of the accompanying radiation
one can apply the well known formulae, and the radiation itself can be considered as
a {\it curvature radiation}.

In the case of large pitch-angles in the curved inhomogeneous magnetic field,
the inhomogeneity does not affect the radiation if the Larmor radius
is small compared to the curvature of the field lines, and to the characteristic distance
scale on which the field is noticeably changed.
This case corresponds to the regime of {\it synchrotron radiation}. All formulae that
describe the synchrotron and curvature radiation regimes coincide if we present them
in the terms of the trajectory curvature radius $R_c$.

For small pitch angles the picture can be essentially different. Let us assume
that initially the velocity vector is parallel to the field line. Then, the acceleration
$\b a\big|_{t=0}=0$, correspondingly the curvature radius\footnote{Note that at the motion with a constant absolute speed value, the curvature radius is equal, by definition, to $R_c=v^2/|\b a|$.} $R_c\big|_{t=0}=\infty$, thus no radiation take place at $t=0$. In the curved magnetic field, a rectilinear uniform motion is not possible, therefore an acceleration and radiation of the particle are unavoidable,
and, what is essential for the correct treatment of the problem, the acceleration and the
curvature radius appear to be time-dependent. In this paper we show that this is true
for all trajectories located close to the {\it smooth trajectory}.
This implies that the radiation also varies with time, thus
for calculations of measurable characteristics of radiation one should average the relevant
distributions over time.

It is generally believed that for the initially small pitch angles the charged particles
rapidly radiate away the fraction of their energy related to the component of the momentum
perpendicular to the field line, i.e. the perpendicular motion quickly disappears, and particles move
along the field lines (see e.g. Ref.~\cite{Usov}).
However, here one should distinguish two different cases. Let
$p_{\parallel}^{}$ and $p_{\perp}^{}$ be the particle momentum components
parallel and perpendicular to the smooth trajectory, respectively. Let us assume that
$p_{\parallel}^{}\gg
mc$ and $p_{\parallel}^{}\gg p_{\perp}^{}$. If $p_{\perp}^{}\ll mc$, then the perpendicular component
of motion is quickly damped; the particle will continue to move along the smooth trajectory and thus
emit curvature radiation. However, if $p_{\perp}^{}\gg mc$, then as in the case of the uniform magnetic field,
$p_{\parallel}^{}$ and $p_{\perp}^{}$ will be reduced with the same time-dependency. The calculations
of characteristics of the resulting radiation, which is different from both synchrotron and curvature
radiation components, is the prime objective of this paper. We call this regime of radiation as
{\it small pitch angle magnetobremsstrahlung}.

\section{The smooth and close to them trajectories}

We start with the analysis of motion of ultrarelativistic particle in a simple model which allows exact
solutions. Namely, below we consider the case when the magnetic field has the same symmetry
as the field of an infinitely long straight wire. Then, in the cylindrical system of coordinates,
$(r,\,\phi,\,z)$, the field has only an azimuthal component; it depends only on the distance
to the axis of symmetry, $r$, but not on $z$ and $\phi$. We denote by $(\b e_r,\b e_\phi,\,\b e_z)$
a moving system of unit vectors linked to the point $(r,\,\phi,\,z)$. The magnetic field can be
expressed in the form $\b B=B(r)\,\b e_{\phi}^{}$. The field lines are circles of radius $r$. It appears that
particles in such a field can move along trajectories close to the circular field lines.
%Let us find these trajectories.

The equation that describes the particle motion in the magnetic field is
\be\label{cur1}
m\gamma\dot{\b v}=\frac ec\,(\b v\times\b B)\,.
\ee
Let us find a solution for which $r=r_0={\rm const}$, and the velocity
\be\label{cur2}
\b v=v_\phi\b e_\phi+v_z \b e_z\, .
\ee
Note that the components $v_\phi$ and $v_z$ are constants. The vector $\b e_\phi$
is uniformly rotating with an angular frequency $\omega=v_\phi/r_0$, therefore the
derivative $d\b e_\phi/dt= -v_\phi\b e_r/r$, and the vector $\b e_z ={\rm const}$.
Substituting Eqs. (\ref{cur2}) into (\ref{cur1}), we find
\be\label{cur3}
m\gamma \frac{v_\phi^2}{r_0}\,\b e_r=\frac ec\,v_z B\,\b e_r\,.
\ee
From here follows that
\be\label{cur4}
\frac{v_z}{v_\phi}=\frac{mcv_\phi\gamma}{eB\,r_0}\,.
\ee
Writing the velocity components in the form $v_z=v\sin\alpha$, $v_\phi=v\cos\alpha$
(the angle starts from the ``equator'', i.e. $-\pi/2\le \alpha \le \pi/2$),
and expressing the velocity $v$ via Lorentz factor, Eq.(\ref{cur4}) can be presented in the following form
\be\label{cur5}
\frac{\sin\alpha}{\cos^2\alpha}=\frac{mc^2\gamma}{eB\,r_0}\,
\sqrt{1-1/\gamma^2
} \, .
\ee
Apparently, for any given $\gamma$ this equation defines the angle $\alpha$.
Eq.(\ref{cur1}) indeed has solutions in the form of Eq.(\ref{cur2}) which describe the
helicoidal motion depending of parameters $\gamma$ and $r_0$. Note that motion of
particles with strictly circular trajectories is not possible.

The condition for a small step in the helix, $|\alpha| \ll 1$, requires
\be\label{cur6}
1\ll \gamma \ll \frac{|e|B\,r_0}{mc^2}=6\times 10^{13} \Big(\frac{B}{10^{11}
\;{\rm G}}\Big) \Big( \frac{r_0}{10\;{\rm km}}\Big)\,.
\ee
Note that typical for pulsars values of $B$ and $r_0$ readily satisfy this condition.
Then,
\be\label{cur7}
\alpha=\frac{mc^2\gamma}{eB\,r_0}\,; \qquad |\alpha|\ll 1 \, ,
\ee
thus the particle trajectory can be interpreted as a motion along the field line with a simultaneous drift
in the perpendicular direction. The positive and negative charges drift in opposite directions.
Under the condition of Eq.(\ref{cur6}), the curvature radius $R_c$ of a particle located at the distance
$r_0$, does not depend on the particle energy and practically coincides with the curvature radius
of the field line $r_0$. Note that this conclusion is correct for any configuration of the
magnetic field. Indeed, the short segments of the field line can be treated as a circle,
therefore when satisfying the condition of Eq.(\ref{cur6}), always exist trajectories which
(or at least their initial segments) are very close (but not identical) to the field line. In the Introduction,
such trajectories have been called smooth trajectories. Also as indicated above, the
magnetobremstrahlung of a particle moving along a smooth trajectory
should be considered as the strict definition of the curvature radiation. Its intensity is defined as
\be\label{cur8}
I_{\rm curv}=\frac{2e^2c}{3}\,\frac{\gamma^4}{R_c^2}
\approx \frac{2e^2c}{3}\,\frac{\gamma^4}{r_0^2}\,.
\ee
When the particle moves close to but not exactly along the smooth trajectories, the
radiation intensity can deviate from Eq.(\ref{cur8}). Therefore one should investigate the
trajectories in the vicinity of the smooth trajectories.

For this purpose it is convenient to use the Hamiltonian formalism. The vector potential can be taken in the form $A_x=0$, $A_y=0$, $A_z=A(r)$, therefore the azimuthal component of the field $B(r)=-dA/dr$.
In the cylindrical coordinates the Hamilton function can be written as
\be\label{cur9}
H=c\,\sqrt{P_r^2+\frac{P_\phi^2}{r^2}+(P_z-\frac ec\,A(r))^2+m^2c^2}\,,
\ee
%ü
where $P_r$, $P_\phi$ and $P_z$ are the generalized momenta corresponding to the coordinates
$r$, $\phi$ and $z$. Because of the azimuthal symmetry of the magnetic field and
its homogeneity along the $z$ axis, $\phi$ and $z$ are cyclic coordinates. Therefore
$P_\phi$ and $P_z$, as well as the energy, are integrals of motion. The presence of three integrals of
motion allows us to reduce the problem to the one-dimensional case, and thus comprehensively
study the qualitative behavior of the solutions. Below we will limit the treatment of the problem
by the case of ultrarelativistic particles and small pitch-angles.

For given $P_\phi$ and $P_z$, the radial motion can be considered, as it follows from Eq.(\ref{cur9}),
as a motion in the field with the following effective potential:
\be\label{cur10}
U_{\rm eff}(r)=\frac{P_\phi^2}{r^2}+(P_z-\frac ec\,A(r))^2\,.
\ee
The first and second derivatives of this potential are
\be\label{cur11}
U'_{\rm eff}(r)=-\frac{2P_\phi^2}{r^3}+\frac{2e}{c}B(r)
(P_z- \frac ec\,A(r))\,,
\ee
and
\be\label{cur12}
U''_{\rm eff}(r)=\frac{6P_\phi^2}{r^4}+\frac{2e}{c}B'(r)
(P_z- \frac ec\,A(r))+\frac{2e^2}{c^2}\,B^2(r)\,.
\ee
Assuming that at $r=r_0$ the derivative $U'_{\rm eff}(r_0)=0$, then at this point
\be\label{cur13}
P_z-\frac{e}{c}A(r_0)=\frac{cP_\phi^2}{r_0^3 e B(r_0)}\,,
\ee
\be\label{cur14}
U''_{\rm eff}(r_0)=\frac{2P_\phi^2}{r_0^4}\left(3+\frac{r_0B'(r_0)}{B(r_0)} \right) +\frac{2e^2}{c^2}\,B^2(r_0)\,.
\ee

If $B$ decreases with $r$ as a power-law, $B\sim 1/r^\d$, then for $\d <3$ the both
terms in Eq.(\ref{cur14}) are positive, therefore $U_{\rm eff}$ has a minimum at $r_0$.
Furthermore, assuming that the condition of Eq.(\ref{cur6}) is fulfilled, in Eq.(\ref{cur14})
we can keep only the last term.

For integration of the equations of motions we apply the standard method which is used
in the theory of small oscillations. In the vicinity of the minimum point the function
$U_{\rm eff}(r)$ is approximated by a parabola:
\be\label{cur15}
U_{\rm eff}(r)=P_0^2+m^2\omega_c^2(r-r_0)^2\,,
\ee
where
\be\label{cur16}
\omega_c^2=\frac{1}{2m^2}U''_{\rm eff}(r_0)\approx
\left(\frac{eB(r_0}{mc} \right)^2\,,
\ee
\be\label{cur17}
P_0^2=U_{\rm eff}(r_0)=
\frac{P_\phi^2}{r_0^2}\left(1+\frac{P_\phi^2}{r_0^4m^2\omega_c^2} \right).
\ee
In these denotations, the Hamilton function can be presented in the form
\be\label{cur18}
H=c\sqrt{P_r^2+m^2\omega_c^2(r-r_0)^2+P_0^2+m^2c^2}\, ,
\ee
and correspondingly the equations of motion are
\be\label{cur19}
\dot r=\frac{\p\tilde H}{\p P_r}=\frac{c^2P_r}{E}\,,\quad
\dot P_r=-\frac{\p\tilde H}{\p r}=-\frac{m^2c^2\omega_c^2}{E}\,(r-r_0)\,,
\ee
where $E$ is the particle energy. The solution of these equations gives
\be\label{cur20}
r=r_0-\rho\cos\frac{\omega_ct}{\gamma}\,,\quad
P_r=m\rho\omega_c\sin\frac{\omega_ct}{\gamma}\,,
\ee
where $\gamma=E/(mc^2)$ is the Lorentz factor of the particle, $\rho$ is
an arbitrary constant\footnote{Note that here the origin of the coordinate system
is arbitrary, therefore instead of $t$ one can use $t-t_0$, where $t_0$ is the
second arbitrary constant. Below we omit the arbitrary constants which
can be added to $\phi$ and $z$.} related to the particle energy through the relation
\be\label{cur20a}
E=c\sqrt{P_0^2+m^2\rho^2\omega_c^2+m^2c^2}\,.
\ee
The condition of applicability of Eq.(\ref{cur20}) is the smallness of $\rho$ compared to $r_0$ as
well as to the characteristic linear scale on which the magnetic field is changed significantly.

Now we should find the time-dependencies of other coordinates.
In this regard we note that
\be\label{cur21}
\dot\phi=\frac{\p H}{\p P_\phi}=\frac{P_\phi}{m\gamma}\,\frac{1}{r^2}
\approx \frac{P_\phi}{m\gamma}\,\frac{1}{r_0^2}
\left(1-\frac{2(r-r_0)}{r_0}\right) \,.
\ee
Using Eq.(\ref{cur20}), we find
\begin{eqnarray}
\dot\phi&=&\frac{P_\phi}{m\gamma}\,\frac{1}{r_0^2}
\left(1+\frac{2\rho}{r_0}\cos\frac{\omega_ct}{\gamma}\right),\label{cur22}\\
\phi&=&\frac{P_\phi}{m\gamma}\,\frac{1}{r_0^2}
\left(t+\frac{2\rho\gamma}{\omega_c r_0}\sin\frac{\omega_ct}{\gamma}\right).\label{cur23}
\end{eqnarray}
Now substituting in
\be\label{cur24}
\dot z=\frac{\p H}{\p P_z}=\frac{c^2}{E}\,\left(P_z-\frac ec A(r)\right)
\ee
the following approximate presentation for $A(r)$,
\be\label{cur24a}
A(r)=A\big(r_0+(r-r_0)\big)\approx A(r_0)-B(r_0)\cdot(r-r_0) \, ,
 \ee
 and using Eqs. (\ref{cur13}) and (\ref{cur20}), we find
\begin{eqnarray}
\dot z&=&\frac{P_\phi^2}{\gamma\omega_c m^2r_0^3}-\frac{\rho\,\omega_c}{\gamma}
\cos\frac{\omega_ct}{\gamma}\,,\label{cur25}\\
z&=&\frac{P_\phi^2}{\gamma\omega_c m^2r_0^3}\,t-\rho
\sin\frac{\omega_ct}{\gamma}\,.\label{cur26}
\end{eqnarray}

In order to make the analytical expressions more convenient to work with,
it is useful to use instead of $P_\phi$ and $\rho$ the following parameters
$\beta_\parallel^{}$ and $\beta_\perp^{}$ which are determined as
\be\label{cur27}
\beta_\parallel^{}= \frac{P_\phi}{mc\gamma r_0}\,,\quad
\beta_\perp^{}=\frac{\rho\,\omega_c}{\gamma c} \, ,
\ee
and denote
\be\label{cur28}
\beta_D^{}=\frac{P_\phi^2}{c\gamma\omega_c m^2r_0^3}=\frac{c\gamma \beta_\parallel^2}
{r_0\,\omega_c}\,.
\ee

Since $P_\phi$ is the projection of the particle momentum
on the axis $z$, then $\beta_\parallel^{}$ is the component of the velocity
parallel to the magnetic field (in units of $c$), and
$\beta_\perp^{}$ is the perpendicular components of the velocity. Finally
$\beta_D^{}$ is the drift velocity.

For the new variables we find
\be\label{cur29a}
 \dot r/c=\beta_\perp^{} \sin\tau\,,\quad \dot z/c=\beta_D^{}-\beta_\perp^{}
\cos\tau\,,
 \ee
 \be\label{cur29b}
\dot\phi= \frac{\beta_\parallel^{}c}{r_0}
\left(1+\frac{2\beta_\perp^{}\beta_D^{}}
{\beta_{\parallel}^2}\,\cos\tau\right),
\ee
where $\tau=\omega_c t/\gamma$; azimuthal component of velocity is
\be\label{cur29c}
v_\phi^{}=r\dot\phi=c\beta_{\parallel}^{}\left(1+\frac{\beta_\perp^{}\beta_D^{}}
{\beta_{\parallel}^2}\,\cos\tau\right).
\ee

For $\beta_\perp^{}=0$, this solution coincides with the obtained above exact solution
that describes the motion along a smooth trajectory (in this case, along the helicoidal line).
The same trajectory can be found if we average Eqs.(\ref{cur29a})-(\ref{cur29c})
over time. Therefore this solution can be interpreted as a motion along the spiral around the smooth trajectory. At $\beta_\perp^{}\ne 0$, the solution is approximate; it is correct when the following condition is fulfilled:
\be\label{cur30}
\beta_\perp^{}\ll \beta_\parallel^{}\,,\qquad
\beta_D^{}\ll \beta_\parallel^{}\,.
\ee
The relation between $\beta_\perp^{}$ and $\beta_D^{}$ can be arbitrary.
Since $\beta_\parallel^{}\approx 1$, the second of the above conditions is equivalent to Eq.(\ref{cur6}).
The same requirement can be formulated in terms smallness of the Larmor radius compared to the
curvature radius of the field line.

Note that the inclusion of the process of the {\it gradient drift} into consideration hardly can be justified since it would exceed the accuracy of the approach. Indeed, the gradient drift would appear in the formulae if one adds in Eq.(\ref{cur24a}) the next (quadratic) term of expansion, while our study is limited
for the linear terms. Therefore throughout the paper we will take into account only the curvature drift.

For the chosen new variables, the velocity and its square are
\be
\b v=\dot r\b e_r+r\dot\phi\b e_\phi+\dot z\b e_z\,,
\ee
and
\be
\b v^2=c^2\left(\beta_\parallel^2+\beta_\perp^2+\beta_D^2 \right)\,.
\ee
Note that $\b v^2$ does not depend on $t$.
In order to find the acceleration, we use the following equation
\be\label{cur31}
\b a=\frac{e}{mc\gamma}\,(\b v\times\b B)=
\frac{\omega_c}{\gamma}\,(\dot r\b e_z-\dot z\b e_r)\,,
\ee
where it is taken into account that the field has only an azimuthal component. The square of the
acceleration
\be\label{cur32}
\b a^2=a_0^2\left(1-\frac{2\beta_\perp^{}}{\beta_D^{}}
\cos\tau+\frac{\beta_\perp^2}{\beta_D^2}\right),
\ee
where $a_0^{}=c^2\beta_\parallel^2/r_0$.
This equation can be derived also
through the consideration of kinematics. Indeed,
\be\label{cur33}
\b a=\ddot{\b r} =(\ddot r-r\dot\phi^2)\b e_r+(2\,\dot r\dot\phi+r\ddot\phi)\b e_\phi+\ddot z\b e_z\,.
\ee
From here, due to the smallness of the ratios $\beta_\perp^{}/\beta_\parallel^{}$ and
$\beta_D^{}/\beta_\parallel^{}$, we obtain an equation identical to Eq.(\ref{cur32}).

At $\beta_\perp^{}=0$ we have $\b a^2=a_0^2={\rm const}$. This is the case
when the particle moves along the smooth trajectory. For
$\beta_\perp^{}=\beta_D^{}$, the acceleration $\b a^2=2\,a_0^2\,(1-\cos\tau)$.
For the moments of time, when the acceleration becomes zero,
the particle velocity is parallel to the field line.
In the limit of $\beta_\perp^{}\gg \beta_D^{}$, the acceleration does not depend on the
radius of the field curvature and on time:
\be\label{cur34}
\b a^2=a_0^2\,\frac{\beta_\perp^2}{\beta_D^2}=
\left(\frac{c\beta_\perp^{}\omega_c}{\gamma}\right)^{\!2}\,.
\ee
It is well known (see e.g. Ref.~\cite{Landau2}) that this equation describes also the acceleration of the particle in the
homogeneous magnetic field. For applicability of Eq.(\ref{cur34}) it is sufficient
to fulfill only the condition $\beta_\perp^{}\gg \beta_D^{}$, while the
relation between $\beta_\perp^{}$ and $\beta_\parallel^{}$ can be arbitrary.

For an ultrarelativistic particle, the curvature radius of trajectory and the acceleration are
linked with a simple relation $|\b a|=c^2/R_c$, which at $v=c$ is, in fact, the definition of the
curvature radius. Thus, the radius of the curvature of trajectory is
\be\label{cur35}
R_c(t)=r_0\left(1-\frac{2\beta_\perp^{}}{\beta_D^{}}
\cos\tau+\frac{\beta_\perp^2}{\beta_D^2}\right)^{\!-1/2}\, .
\ee
Here $\beta_\parallel$ is set to be equal to 1 given that
$1-\beta_\parallel^{}\ll 1$.

Let us consider now the motion of the charged particle in the field of a
magnetic dipole (the so-called St\"ormer problem). Such a system
appears to be (unlike the previous case) non-integrable (see e.g. Ref.~\cite{Dragt}),
and generally it is impossible to conduct even a qualitative analysis of the motion.
However, when the condition of Eq.(\ref{cur6}) is fulfilled, the system has smooth trajectories.
The trajectories in the close vicinity of the latter present lines spiraling around the smooth trajectories,
i.e. basically we deal with the same picture as discussed above for the straight wire. This is
confirmed also by our numerical calculations.

\begin{figure}
\begin{center}
\includegraphics[width=0.33\textwidth,angle=-90]{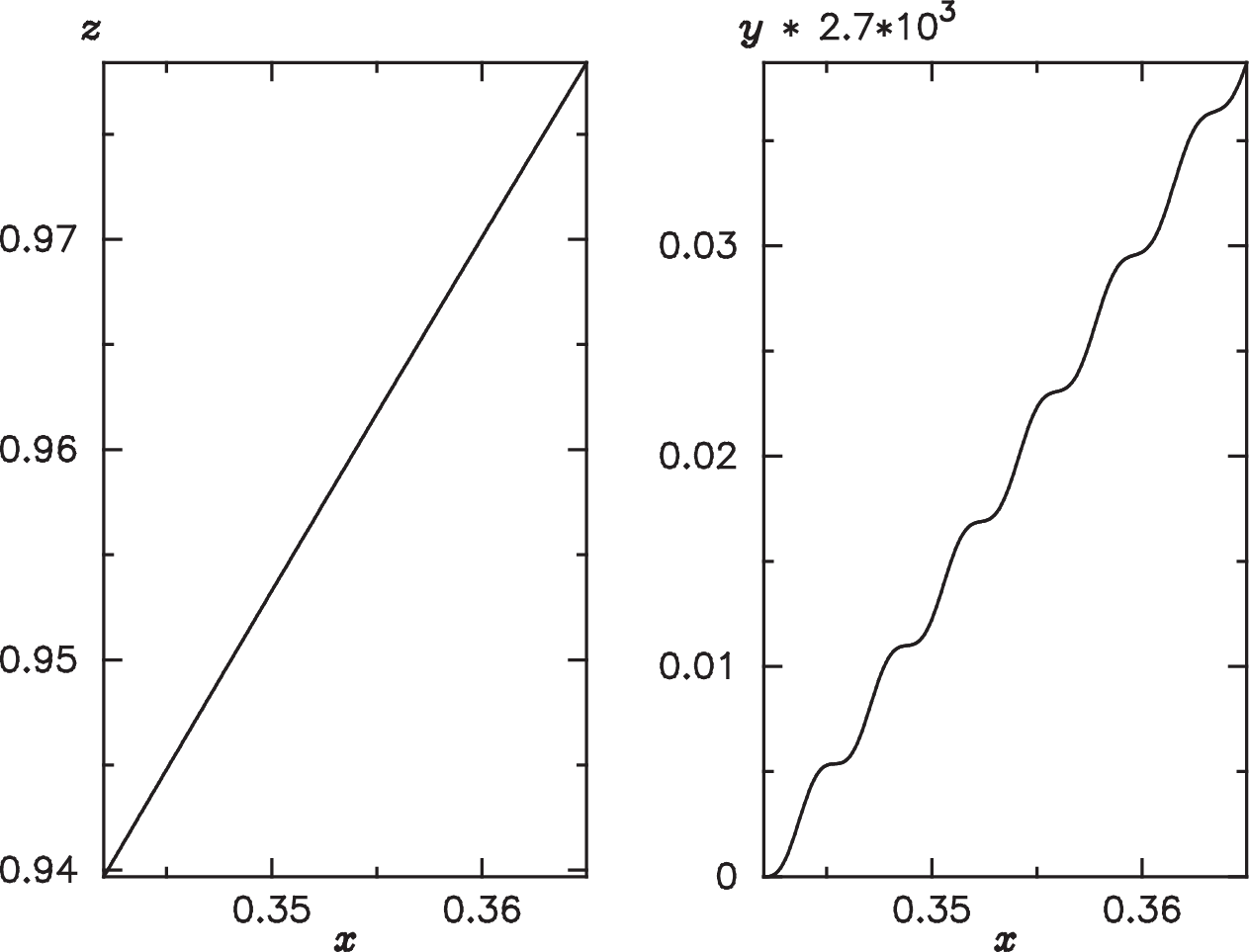}
\caption{\small The initial segment of the trajectory of charged particle in the field of the magnetic dipole.
The initial point is located on the surface of the star at the angle $20^\circ$ relative to the $z$-axis, the
initial velocity is directed along the field line. The y-axis in the right panes is rescaled (multiplied by a factor of $2.7 \times 10^3$ in order to make visible the oscillations.}
\label{zxy}
\end{center}
\end{figure}

\begin{figure}
\begin{center}
\includegraphics[width=0.33\textwidth,angle=-90]{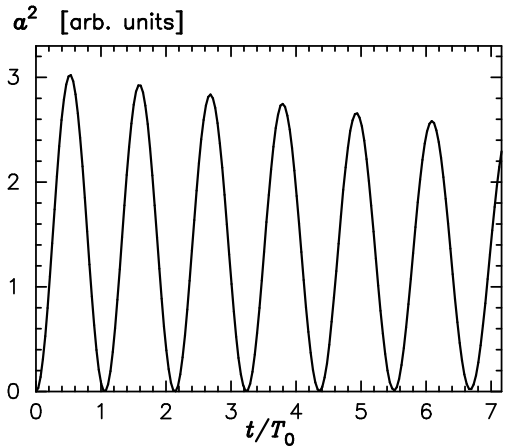}
\caption{\small The time-dependence of acceleration during the motion of a
charged particle along the trajectory presented in Fig.\ref{zxy}
\label{ta}}
\end{center}
\end{figure}

In Fig.~\ref{zxy} we show the initial part of the trajectory in the field of magnetic
dipole directed towards the $z$-axis. At the initial moment of time the particle is located
in the point $\b r(0)=(\sin\vt,0,\cos\vt)$ where the coordinates are measured in units of the radius
of the object (star), $R_*$, the angle $\vt =20^\circ$, the initial velocity is directed along the field line, and the parameter $\xi=10^{-3}$ (see Appendix~\ref{app}). The amplitude of oscillations appears
of the same order of magnitude, i.e.
$10^{-3}$, therefore in the left panel of Fig.~\ref{zxy} the oscillations are not visible.
Fig.~\ref{ta} shows the dependence of acceleration on time. The period of oscillations is
equal to $T_0=2\pi mc\gamma/(eB_0)$, where $B_0$ is the magnetic field on the surface of the star.
The decrease of the amplitude and the increase of the period is due to the reduction of the field
along the trajectory. The change of the Lorentz factor $\gamma$ due to the radiation
losses is not taken into account. The trajectory patters for larger time intervals
are discussed in Appendix.

%FIG 1

%FIG 2

It should be noted that the necessity of calculations of characteristics of (synchro-bremsstrahlung) radiation
in the context of proper treatment of trajectories of charged particles in curved magnetic fields, has
been correctly indicated also in Ref.~\cite{Cheng}. However, our analysis shows that some of the
basic results and conclusions of that paper appear to be not correct. Therefore for clarification of
some issues of conceptual character, below we compare the results of the present study
with the relevant result from Ref.~\cite{Cheng}. For convenience,
the equations from Ref.~\cite{Cheng} will be referred by their original numbers with an additional
indication of \cite{Cheng}.

Let us compare, in particular, the results of calculations regarding
the square of the acceleration, $\b a^2$,
in the curved magnetic field.
For the denotations used in this paper, the equation (5,\cite{Cheng}) can be written in the form
\be\label{cheng1}
\b a^2=a_0^2\left(1+\frac{\beta_\perp}{\beta_D}\right)^{\!2} \, .
\ee
This equation is not correct; in particular it lacks time-dependence. Apparently, the error in
Ref.~\cite{Cheng} has appeared because of the definition of acceleration for which the authors
used a wrong formula given by equation (1,\cite{Cheng}). The reference to the monograph
of Jackson \cite{Jackson}, as an original source of this equation is misleading.
The monograph \cite{Jackson} does contain derivation of the drift velocity. The second term in
(1,\cite{Cheng}) in fact is the drift velocity $\b v_D^{}$ (Eq.(12.117) from Ref.~\cite{Jackson}).
In the derivation of $\b v_D^{}$ a time-averaging procedure was applied, thus
all oscillating terms have been canceled. If we are interested only in the systematic shift (drift),
these terms, of course, do not affect the result. However if one wants to calculate the radiation of particles, then the oscillating terms cannot be ignored. Correspondingly, the velocity $\b v(t)$ cannot be obtained by adding to $\b v_D^{}$ a constant vector (of unknown direction!) $\b v_0$, as it is done in
Ref.~\cite{Cheng}.

It should be noted that in some cases the replacement of $\b v(t)$ by a constant vector
should not necessarily lead to a wrong results. For example, from the solution of the
equation for the motion in the constant homogeneous
field follows that $\b a^2={\rm const}$, therefore for definition of $\b a^2$, we can ignore
the velocity's time-dependence. We would like to emphasize that one can make such
a statement only when the solution of the equation motion is known. In more complex cases,
like the one analyzed above, the solution of the equation of motion is a necessary step.
Any {\it a priori} assumption that $\b a^2(t)={\rm const}$ may lead to incorrect results.
In this regard, all equations in Ref.\cite{Cheng}, which contain $\b a^2$ are not correct.
The results of this work can be considered, in the best case, as interpolations
of two well known cases: $\beta_\perp^{}\ll \beta_D^{}$ (curvature radiation) and
$\beta_\perp^{}\gg\beta_D^{}$ (synchrotron radiation). In these limiting cases the results of Eqs.~(\ref{cheng1}) and (\ref{cur32}) coincide. The reason of this agreement is explained above.

Finally, we note that Ref.~\cite{Cheng} contains another incorrectness related to the derivation of Eq.(21,\cite{Cheng}). Although the authors do cite to the classical paper of J.~Schwinger \cite{Schwinger}, apparently they have missed to mention that not only the final result (i.e. the equation (21,\cite{Cheng}) itself), but also all intermediate steps for derivation of this equation (using the same method and the same denotations)
have been already obtained in Schwinger's paper.

\section{Intensity and energy spectrum of radiation}

In this section we calculate, within the framework of classical electrodynamics,
the intensity and the energy spectrum of radiation of a charged ultrarelativistic
particle with the perpendicular component of velocity $\beta_\perp\sim\beta_D$.
If the radius of the curvature of the trajectory $R_c\sim r_0$, the characteristic energy of the
emitted photon is $\sim \hbar c\gamma^3/r_0$. The requirement of smallness of this energy
compared to the energy of the particle $mc^2\gamma$ (the condition of applicability
of classical electrodynamics) implies $\hbar\gamma^2\ll mcr_0$. We assume that
the perpendicular (relative to the smooth trajectory) momentum $p_\perp^{}\ll p_\parallel^{}$, but,
at the same time, $p_\perp^{}\gg mc$. Using Eq.(\ref{cur28}), the latter inequality can be presented in the
form $c\gamma^2\gg \omega_c r_0$. This gives the following constraint on the Lorentz-factor:

\be\label{cond}
\omega_c r_0/c \ll \gamma^2 \ll mcr_0/\hbar\,.
\ee
The condition of smallness of the lower limit compared to the upper limit, implies
$\hbar\omega_c\ll mc^2$, i.e. the magnetic field should not exceed the critical one, $B_{\rm cr}=m^2c^3/(e\hbar)=4.4\times 10^{13}\;{\rm G}$.

The energy radiated away by the particle for a unite interval of time, is
\cite{Landau2}
\be\label{cur36}
I=-\frac{dE}{dt}=\frac{2e^2\gamma^6}{3c^3}\,(\b a^2-(\b v\times \b a)^2/c^2 )
=\frac{2e^2c}{3}\,\frac{\gamma^4}{R_c^2}\,.
\ee
Substituting $R_c$ from Eq.(\ref{cur35}) , we obtain
\be\label{cur37}
I=\frac{2e^2c}{3}\,\frac{\gamma^4}{r_0^2} \left(1-2\eta\cos\tau+\eta^2\right)\,,
\ee
where we introduce a new parameter $\eta=\beta_\perp^{}/\beta_D^{}$.
The averaged over the period intensity is
\be\label{cur38}
\langle I\rangle=\frac{2e^2c}{3}\,\frac{\gamma^4}{r_0^2} \left(1+\eta^2\right).
\ee
This formula can be written in the form $\langle I\rangle=I_{\rm curv}(1+\eta^2)$,
where $I_{\rm curv}$ is defined in Eq.(\ref{cur8}). It is seen that during the motion in the
curved magnetic field the curvature radiation has the minimum possible intensity.

In the considered scenario, we deal with three characteristic times. These are the time of energy losses
(cooling time), $t_{\rm cool}\sim mcr_0^2/(e^2\gamma^3)$, the period of ``oscillations",
$T\sim \gamma/\omega_c$, and the characteristic time of formation of radiation,
 $\Delta t\sim r_0/(c\gamma)$. It is easy to show, using Eq.(\ref{cond}) , that $\Delta t\ll t_{\rm cool}$.
 Note that this condition is always satisfied in the framework of classical electrodynamics.
 The ratio $\Delta t/T\sim \omega_c r_0/(c\gamma^2)$ also is small compared to 1,
as it follows from Eq.(\ref{cond}). Finally, the ratio
$T/t_{\rm cool} \sim e^2\gamma^4/(mc\omega_cr_0^2)$; using the lower and upper limits of
the Lorentz-factor from Eq.(\ref{cond}), we obtain
\be\label{cond1}
\alpha\,\frac{\hbar\omega_c}{mc^2}\ll \frac{T}{t_{\rm cool}} \ll
 \alpha\,\frac{mc^2}{\hbar\omega_c}\,,
\ee
where $\alpha=e^2/\hbar c\approx 1/137$.

The smallness of $\Delta t$ simplifies significantly the calculations of the radiation spectrum.
This allows us to ignore the changes of the particle energy and the curvature radius with time, and
perform calculations at the fixed "prompt" values of these parameters:
$E(t)=mc^2 \gamma(t)$ and $R_c(t)$. Then the spectral flux density of the magnetobremsstrahlung
integrated over the all emission angles, is described by the well known
expression for the synchrotron regime of radiation (see, e.g. Refs.~\cite{Schwinger}, \cite{Ginzburg})
\be\label{cur39}
P(\omega,t)=\frac{\sqrt{3}\,e^2}{2\,\pi}\,\frac{\gamma}{R_c}
\,F\!\left(\frac{\omega}{\Omega_*}\right).
\ee
Here
\be\label{cur40}
\Omega_*=\frac{3c\gamma^3}{2R_c}\,,
\ee
and
\be\label{cur41}
F(x)=x\int_x^\infty\! K_{5/3}(x')\,dx'\,,
\ee
where $K_{5/3}$ is modified Bessel function. The particle energy loss rate,
$dE/dt=-\int_0^\infty\!P(\omega,t)\,d\omega$; the calculation of this integral
results, as expected, in Eq.(\ref{cur36})

Since $R_c$ and $\gamma$ vary with time, the radiation spectrum should be time-dependent as well.
Let assume that the condition $T\ll t_{\rm cool}$ is satisfied (note that this condition does not contradict to
Eq.(\ref{cond1})). Then, since change of $\gamma$ during the period is small, we can introduce new parameters averaged over time. This in fact has been already done at the changeover from Eq.(\ref{cur37}) to
Eq.(\ref{cur38}), assuming that during the period $\gamma = {\rm const}$.

By writing $R_c$, in accordance with Eq.(\ref{cur35}), as
\be\label{cur42}
R_c=r_0/q(\eta,\tau)\,,\;\quad q(\eta,\tau)=\sqrt{1-2\eta\cos\tau+\eta^2}\,,
\ee
the averaged spectral flux density can be presented in the form
\be\label{cur43}
\langle P(\omega,t)\rangle=\frac{\sqrt{3}\,e^2}{2\,\pi}\,\frac{\gamma}{r_0}
\,G\!\left(\frac{\omega}{\omega_*}\right).
\ee
Here
\be\label{cur44}
G\!\left(\frac{\omega}{\omega_*}\right)=\frac{1}{\pi}\int_0^\pi\!
q(\eta,\tau)\,F\!\left(\frac{\omega}{\omega_* q(\eta,\tau)}\right) d\tau\,,
\ee
where $\omega_*=3c\gamma^3/(2r_0)$. $G$ is a function of two variables: $x=\omega/\omega_*$ and $\eta=\beta_\perp^{}/\beta_D^{}$. If in Eq.(\ref{cur44}) we represent the function $F(x)$
in the form of Eq.(\ref{cur41}), and change the order of integration, the integral over $d\tau$ can be calculated analytically. This allows us to express the function $G(x)$ in the form of a single integral:
\be
G(x)=x\int_{x/(1+\eta)}^\infty\! K_{5/3}(x')\,\Psi(x/x')\,dx'\,,
\ee
where
\be
\Psi\!\left(\frac{x}{x'}\right)=\left\{
\begin{array}{ll}
\frac{1}{\pi}\arccos\frac{(x/x')^2-1-\eta^2}{2\eta}\,, & \frac{x}{x'}\ge |1-\eta|\,,\\
1\,, & \frac{x}{x'}< |1-\eta|\,.
\end{array}
\right.
\ee

\begin{figure}
\begin{center}
\includegraphics[width=0.33\textwidth,angle=-90]{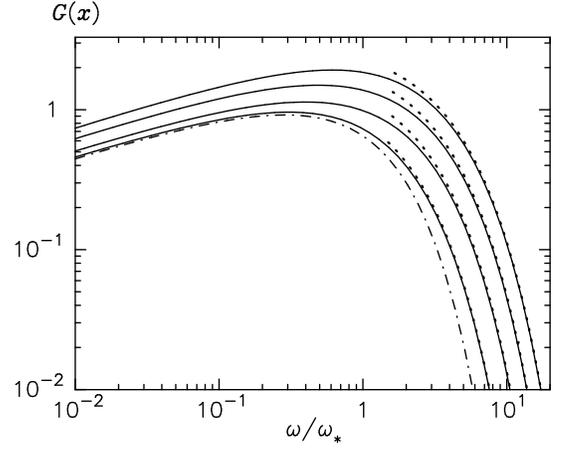}
\caption{\small The spectral flux density of the small pitch-angle radiation averaged over
time. The thin solid lines (from top to bottom) correspond to numerical calculations performed for
the values of the ratio $\eta=\beta_\perp^{}/\beta_D^{}$ = 0, 0.5, 1, 1.5 and 2.
The dotted lines are calculated using the asymptotic presentation of the function $G(X)$ given by
Eq.(\ref{asymp}).
The dot-dashed line corresponds to the curvature radiation ($\eta=0$)
when the particle moves along the soft
trajectory.
\label{f1}}
\end{center}
\end{figure}

The function $G$ for different values of $\eta$ is shown in Fig.~\ref{f1}. At
$\eta=0$ (the dot-dashed line) we have a nominal curvature radiation.
The curves are based on numerical integration of Eq.(\ref{cur44}),
using for the function $F(x)$ the following analytical approximation from Ref.~\cite{Aharonian1},
\begin{eqnarray}
F(x)\approx 2.15\,x^{1/3}\,(1+ 3.06\,x)^{1/6}_{}\nonumber \\
\times \frac{1+0.884\,x^{2/3}+0.471\,x^{4/3}}{1+1.64\,x^{2/3}+0.974\,x^{4/3}}\,
e^{-x}\label{sy_un10}\,,
\end{eqnarray}
which provides an accuracy better than 0.2\% over the entire range of the variable $x$.
In two limits, $x \gg 1$ and $x \ll 1$, the function $F(x)$ can be expressed as
\begin{eqnarray}
F(x)\approx 2^{2/3}\Gamma\!
\left(\frac23\right)x^{1/3}\,,& \quad x\ll 1\,,\label{sy_as1}\\
F(x)\approx \sqrt{\frac{\pi x}{2}}\,e^{-x}\,,&
\quad x\gg 1\label{sy_as2}\,.
\end{eqnarray}
Therefore at $x\ll 1$, the function $G(x)$ differs from
$F(x)$ only by a factor which does not depend on $x$:
\be
G(x)=\frac{1}{\pi}\int_0^\pi\!(1-2\eta\cos\tau+\eta^2)^{1/3}\,d\tau\times F(x)\,.
\ee
In the opposite limit, $x\gg 1$, the integral can be calculated by using the standard
saddle point method. This gives the following asymptotics:
\be\label{asymp}
G\approx \frac{(1+\eta)^2}{2\sqrt{\eta}}\exp\!\left(-\frac{x}{1+\eta}\right)\,.
\ee
Formally, in this case the saddle point method can be applied if
$x\gg(1+\eta)^3/\eta$. However, the numerical calculations show that in the interval
$0.2\le \eta \le 3$ the simple analytical function given by
Eq.(\ref{asymp}) provides a good accuracy (better than 10 \%)
already at $x\approx 2$ (see Fig.~\ref{f1}).

%FIG3

Let's consider now the case when the injected accelerated electrons are not strictly unidirectional
but are distributed within a narrow cone with an opening angle $\Delta\theta\ll 1$.
We assume that all electrons have the same energy. For the standard synchrotron radiation,
a slight change of the pitch angle does not affect the spectrum of radiation.
However, in the case of the small pitch-angle magnetobremsstrahlung, even a tiny
spread of pitch-angles may result in a significant change of the spectrum.
It is convenient to express the angular distribution of particles
through the parameter $\eta$, which is linked to the
pitch angle with a simple relation: $\theta=\eta\beta_D^{}/\beta_\parallel^{}$. Note that at
$\eta\lesssim 1$, the pitch angle $\theta\ll 1$. For calculations we should specify the
angular distribution of particles. Below we consider a
Gaussain type distribution:
\be\label{dis1}
g(\eta)\,d\eta \sim e^{-\eta^2/(2\zeta)}\, d\eta\,,
\ee
thus the mean value of the square of the perpendicular velocity
\be\label{dis2}
\langle\beta_\perp^2\rangle= \zeta\,\beta_D^2\,.
\ee
For determination of the energy spectrum we have to average Eq.(\ref{cur44}) over time $\tau$
and parameter $\eta$, i.e. calculate the following integral
\be\label{cur45}
\langle G(x)\rangle=\int_0^\infty \! d\eta\,g(\eta)\int_0^\pi\!\frac{d\tau}{\pi}
q(\eta,\tau)\,F\!\left(\frac{x}{q(\eta,\tau)}\right).
\ee

\begin{figure}
\begin{center}
\includegraphics[width=0.33\textwidth,angle=-90]{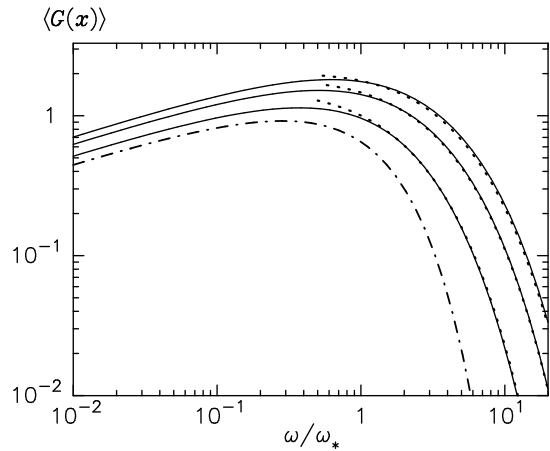}
\caption{\small The spectral flux density of the small-pitch angle radiation averaged over the period and the pitch-angles of particles, assuming a Gaussian type of distribution given by Eq.(\ref{dis1}). The solid lines correspond to different values of
$\zeta=\langle\beta_\perp^2\rangle/\beta_D^2$: 1, 3 and 5 (from bottom to top).
The dotted lines are calculated using the asymptotic analytical presentation given by Eq.(\ref{dis3}).
The dash-dotted line corresponds to the curvature radiation ($\zeta=0$).
\label{f2}}
\end{center}
\end{figure}

The energy spectra calculated for several values of $\zeta$ are shown in Fig.~\ref{f2}.
Apparently, the case of $\zeta=0$ corresponds to the curvature radiation.

As above, using the saddle point method, we can find the
asymptotics for $\langle G\rangle$ at large $x$.
Rather cumbersome calculations, which we omit here, leads to the following result
\be\label{dis3}
 \langle G(x)\rangle\approx\sqrt{\frac{\zeta}{3x^{1/3}}} \left(x^{1/3}+\zeta^{-1/3} \right)^2\,e^{-u}\,,
\ee
where
\be\label{dis4}
 u=\frac32\,\zeta^{-1/3}x^{2/3}-\zeta^{-2/3}x^{1/3}+\frac1{3\zeta}\,.
\ee
This equation is derived under the following conditions:
$x\gg \zeta$ and $x\gg 1/\zeta$. However, the comparison of Eq.(\ref{dis4}) with
accurate numerical calculations show (see
Fig.~\ref{f2}), that for $\zeta\sim 1$ this analytical presentation gives
correct results already at $x\gtrsim 0.5$.

\begin{figure}
\begin{center}
\includegraphics[width=0.33\textwidth,angle=-90]{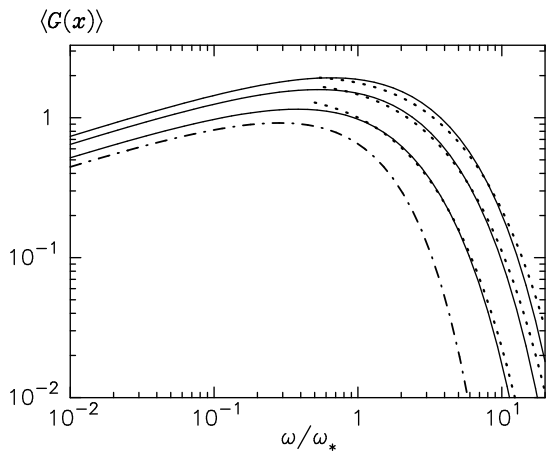}
\caption{\small The spectral flux density of the small-pitch angle radiation averaged over the period and the pitch-angles of particles, assuming a uniform angular distribution of electrons within the opening angle of the beam. The solid lines correspond to different values of the mean square of the perpendicular velocity
$\zeta=\langle\beta_\perp^2\rangle/\beta_D^2$: 1, 3 and 5 (from bottom to top).
The dotted lines are the same as in Fig\ref{f2}. They are calculated for the Gaussian type of angular distribution using the asymptotic analytical presentation given by Eq.(\ref{dis3}).
The dash-dotted line corresponds to the curvature radiation ($\zeta=0$).
\label{flat}}
\end{center}
\end{figure}

Although the assumed Gaussian type angular distribution of electrons seems to be a quite reasonable and natural choice in the considered scenario, it would be interesting to
investigate the dependence of the radiation spectrum on the
specific angular distribution of the electron beam. In particular in Fig.\ref{flat}
we show the radiation spectra calculated for the uniform distribution of electrons
within fixed opening angles of the beam. The results shown in Fig.\ref{flat} (solid curves)
are obtained for the same values of the mean square of the perpendicular velocity as in
Fig.~\ref{f2}. In Fig.\ref{flat} we show also the asymptotic solutions given by Eq.(\ref{dis3})
for the spectra calculated for the Gaussian type angular distribution (the same dotted curves
shown in Fig.~\ref{f2}). The comparison of the results shown in Figs.\ref{f2} and \ref{flat}
indicates that the value of the mean square of the perpendicular velocity
generally characterizes the energy spectrum of radiation
independent of the details of the specific small pitch-angle distribution of electrons.

Note that the spectra of the small-pitch angle radiation do not contain the characteristic
exponential term $e^{-x}$, as in the case of the
synchrotron or curvature radiation, but show a harder behavior. This can be seen from
the asymptotics Eq.(\ref{dis4}). To demonstrate the tendency of steepening of the energy spectrum
beyond the maximum at $x \sim 1$, in the table below we show the evolution of
the local slope of the spectral flux density ( the so-called spectral index $\alpha$)
which is determined as
\be
\alpha=-d(\log \langle G\rangle)/d(\log\omega)\,.
\ee
It is seen that while in the case of curvature radiation ($\zeta=0$), the spectrum becomes very steep already at $x\gtrsim 3$,
 in the small-pitch angle radiation regime with $\zeta =2$ or $\zeta =3$ , a relatively hard spectrum can extend up to $x\sim 10$.

%%%%%%%%%%%%%  TABULAR  %%%%%%%%%%%%%
\begin{picture}(200,90)
\put(5,30){%
\begin{tabular}{|c||cccccc|}
\hline
&2.0 & 3.0 &5.0 & 10. & 20. &30.\\
\hline
\hline
0 &1.65 &2.62 &4.59 & 9.56 & 19.5 & 29.5\\
\hline
\hspace{5pt}0.5\hspace{5pt}&\hspace{5pt}1.17\hspace{5pt}&\hspace{5pt}1.67\hspace{5pt} &\hspace{5pt} 2.56\hspace{5pt} &\hspace{5pt} 4.45\hspace{5pt} & \hspace{5pt}7.56\hspace{5pt} &\hspace{5pt}10.2\; \hspace{8pt}\\
\hline
1.0 &0.96 &1.37& 2.10 & 3.63 & 6.14 & 8.28\\
\hline
2.0 & 0.75 & 1.08 &1.67 & 2.91 & 4.93 & 6.65\\
\hline
3.0 & 0.64&0.93 &1.45 & 2.54& 4.31& 5.82\\
\hline
\end{tabular}
}
\put(5,75){\line(5,-3){22}}
\put(10,61){${}^\zeta$}
\put(19,65){${}^x$}
\end{picture}
%%%%%%%%%%%%%%%%%%%%%%%%%%%%%%%%%%%%%

%FIG4

%FIG5

\section{Discussion}

In the context of the motion of an ultrarelativistic charged particle in the strong magnetic field with
curved field lines, in the vicinity of the latter always exist trajectories without fast oscillations
of the curvature radius. We call them {\it smooth trajectories}. The angle between the smooth
trajectory and the field line appears equal to $\beta_D^{}$ and is determined by Eq.(\ref{cur28}).
For the conditions discussed in this paper, $\beta_D^{}\ll 1$. This trajectory is distinct in the sense that, strictly speaking, the curvature radiation takes place when the charged particle moves along this trajectory, but not along the magnetic field line as it is broadly accepted and discussed in the literature. In fact,
if the velocity vector is initially parallel to the magnetic field line, no radiation is possible {\it by definition}!
The curvature radiation appears when the velocity vector becomes parallel to the smooth trajectory.
On the other hand, since the curvature radius of the smooth trajectory $R_c$ is very close to the
radius of the field curvature $r_0$, the intensity and spectral distribution of the curvature radiation are nevertheless correctly described by the formalism applied in the original papers (see e.g. \cite{Usov}) where the curvature radiation has been linked
to the motion of particles along the magnetic field lines.

Another common misconception in the literature is connected with the belief that the radiation
in the strong magnetic field results in prompt damping of the perpendicular component of motion.
However, if the perpendicular momentum is ultrarelativistic, the parallel and perpendicular
components of the momentum decrease (due to the radiation) by the same law. A strict proof of this
statement is possible in the case of homogeneous magnetic field. However, in the context of
consideration of direction of the recoil momentum, this feature can be extended to the curved
magnetic field for the particle motion along the trajectories close to the smooth trajectory.
The fast damping of the perpendicular motion occurs only if the perpendicular component of
momentum is non-relativistic. This circumstance might have important practical implications
for radiation of highly magnetized relativistic objects.

If initially the velocity vector of the particle is not directed strictly along the smooth trajectory, but
at a small angle ($\theta\ll 1$), then for $\theta\gamma\gg 1$ the perpendicular momentum is
ultrarelativistic and the particle moves along the line spiraling around the smooth trajectory
(but not the field line). In particular, this happens when initially the velocity
is parallel to the field line, provided that $\gamma\beta_D^{}\gg 1$. Then, at
the initial moment of time $\b a=0$ and $R_c=\infty$. The important difference from the
motion in the homogeneous magnetic field is the time-dependence of
the curvature radius and the square of acceleration. The radiation in this regime
is principally different from both the synchrotron and the curvature radiation regimes.
This new regime of the magnetobremsstrahlung, which surprisingly has been missed
in the previous studies, would be appropriate to call {\it small pitch-angle} radiation.

The averaged over the period intensity of radiation is given by Eq.(\ref{cur38}). It can be presented in the standard form of Eq.(\ref{cur36}), if we introduce an {\it effective} curvature radius $R_{\rm eff}=r_0/\sqrt{1+\eta^2}$. However, if we want to have a standard exponential term $e^{-x}$ in an explicit form (as in the case of the synchrotron or curvature radiation) in the asymptotic presentation for the spectrum given by Eq.(\ref{asymp}), the effective curvature radius should be defined differently, namely $R_{\rm eff}=r_0/(1+\eta)$. This is another indication of the difference between the small-pitch angle radiation from the synchrotron and curvature regimes of radiation.

The characteristic feature of the small-pitch angle radiation is its
strong dependence on the particle angular distribution.
Even in the case of a very narrow angular distribution, the
radiation spectrum could differ significantly from the unidirectional beam. This is demonstrated
in Fig.\ref{f2} and \ref{flat} where the radiation spectra averaged over
the pitch-angles of particles, assuming Gaussian type and uniform angular
distributions of particles are presented. For the Gaussian type angular distribution,
the spectral flux density
can be presented in a simple analytical form by Eq.(\ref{dis3}) which
provides a good accuracy already at $x \gtrsim 0.5$. It is seen that the small
pitch-angle radiation predicts
significantly harder energy spectra compared to the curvature radiation
($\zeta=0$).

This interesting feature might  be a key for
interpretation of the recent observations of pulsars by
 {\it Fermi} LAT which indicate that the energy spectra of some pulsars, in
particular the Crab pulsar \cite{Fermi}, in the cutoff region are
significantly harder than $e^{-x}$ as predicted by the curvature radiation
models. The energy spectrum of the pulsed emission of the Crab reported in
Ref.~\cite{Fermi} can be readily described by Eq.(\ref{dis3}) assuming
$\zeta=1$ and a quite reasonable value of the ratio
$\gamma^3/r_0 \simeq 10^{13} \;{\rm cm}^{-1}$.

We should note, however, that the formalism developed in this paper cannot be
directly used for interpretation of experimental data. It can be applied only at
the initial stages, when the energy losses of particles are negligible.
Otherwise, for a correct treatment of the time-evolution of particles and their
time-dependent radiation, one should solve kinetic equations that describe the
energy and space distributions of particles taking into account the particle
energy losses, changes of their trajectories, etc. This issue we will discuss
elsewhere.

Finally, we note that the results of this work are obtained within the framework of classical electrodynamics
when the quantum effects can be ignored. If the condition of Eq.(\ref{cond}) is violated, then
calculations should be conducted using the methods of quantum theory. If in the initial state the electrons
occupies the first level of Landau, then the standard method of calculations used in the
quantum theory of synchrotron radiation (see e.g. Ref.~\cite{Berest}) is not
applicable since the perpendicular motion cannot be treated as quasi-classical.

\appendix

\section{Trajectories in the field of magnetic dipole\label{app}}

% FIG 6
\begin{figure}
\begin{center}
\includegraphics[width=0.33\textwidth,angle=-90]{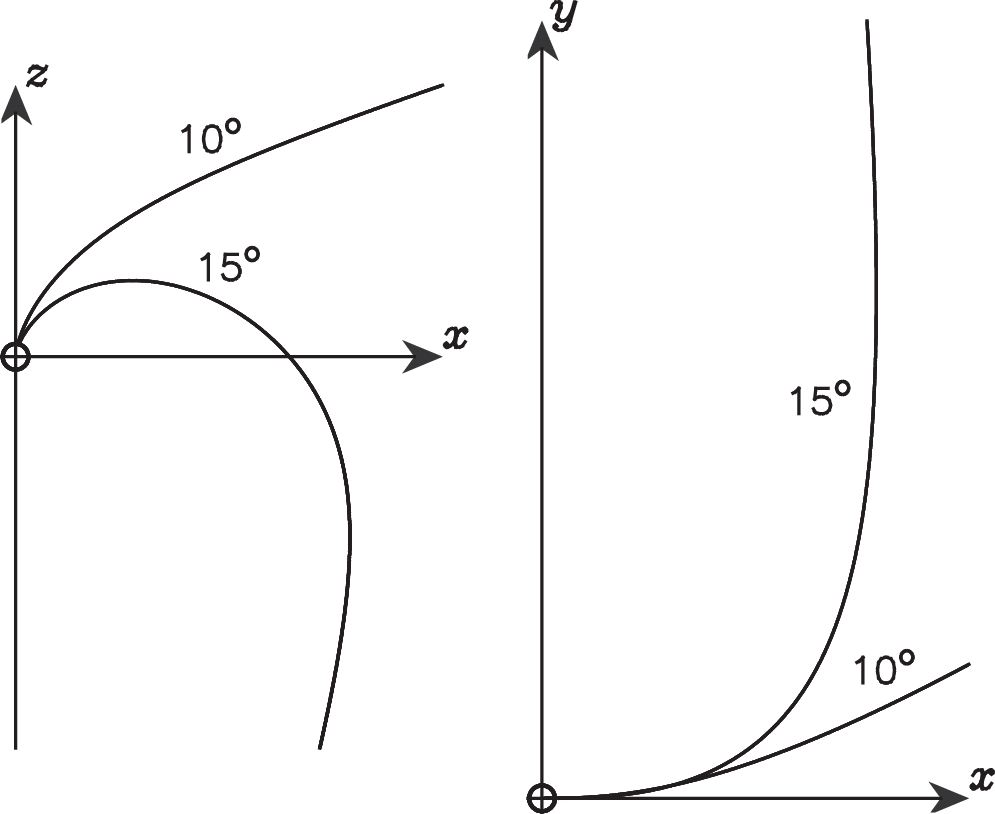}
\caption{\small \label{xyz10} The particle trajectories on the planes
$(x,z)$ (left panel) and $(x,y)$ (right panel) for angles $\vt=10^\circ$ and
$15^\circ$. The motion is infinite, the direction of the final velocity depends on the initial angle
$\vt$. The circle at the origin of coordinates is the surface of the ``star''.}
\end{center}
\end{figure}

Here we present for illustration the results of numerical calculations of trajectories in the field of the magnetic dipole. The initial Cartesian coordinates of the particle (in units of $R_*$) we set on the sphere of radius $R_*$ in the form $\b r(0)=(\sin\vt,\,0,\,\cos\vt)$; the initial velocity is directed along the field line,
which starts from the point $\b r(0)$. It is convenient to introduce, instead of $t$, a dimensionless
parameter $\tau=t\,eB_0/(nc\gamma)$, where $B_0$ is the field in the initial point of the trajectory, and
write down the equation of motion in the following form
\be\label{app1}
\frac{d\nu}{d\tau}=\b\nu\times \b b\,,\quad \frac{d\b
r}{d\tau}=\xi\b\nu\,.
\ee
Here $\b\nu=\b\beta/\beta$ is the unite vector in the direction of velocity, $\b b=\b B(\b r)/B_0$,
\be\label{app2}
\xi= \frac{mc^2\gamma\beta}{eB_0R_*}\,.
\ee
Note that only one parameter, $\xi$, enters into this equation. In accordance with
the condition of Eq.(\ref{cur6}), the values of $1-\beta$ and $\xi$ are small compared to unity.
We adopted $\xi=10^{-3}$, and performed the calculations without taking into account the effect of
radiative friction.

% FIG 7
\begin{figure}
\begin{center}
\includegraphics[width=0.33\textwidth,angle=0]{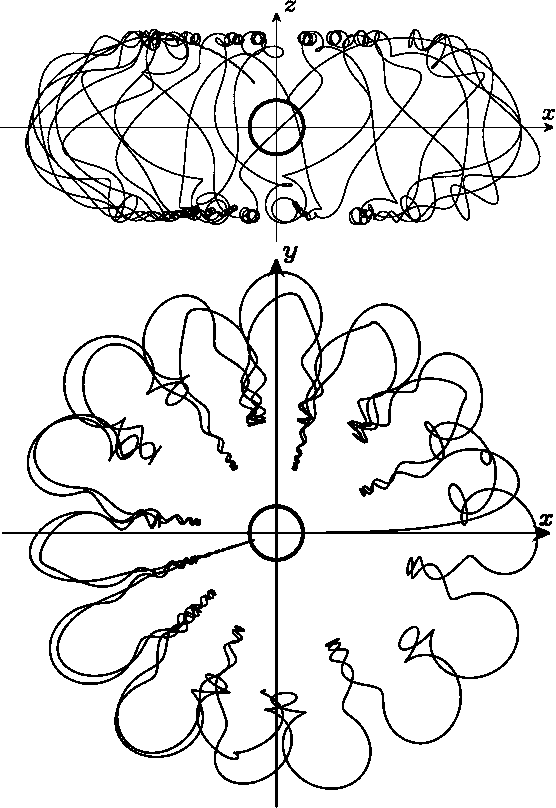}
\caption{\small The same as in Fig.~\ref{xyz10}, but for angle $\vt=20^\circ$.
\label{xyz20}}
\end{center}
\end{figure}

% FIG 8
\begin{figure}
\begin{center}
\includegraphics[width=0.38\textwidth,angle=-90]{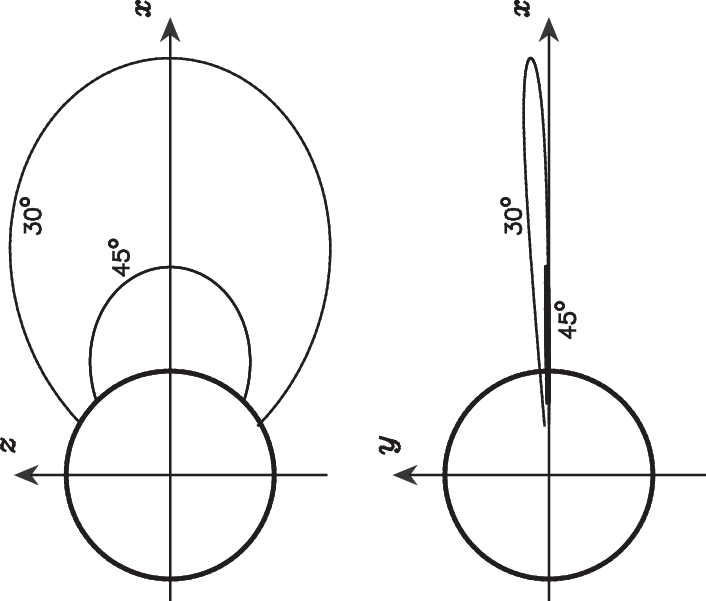}
\caption{\small The same as in Fig.~\ref{xyz10}, but for angles $\vt=30^\circ$
and $45^\circ$. \label{xyz30}}
\end{center}
\end{figure}

If initially we put the particle on the ``pole" and direct the velocity towards the axis $z$,
the particle apparently will move uniformly along that axis. If at $t=0$ the particle is located close to the ``pole", the motion will be not rectilinear, but it will remain infinite (see Fig.~\ref{xyz10}).

With an increase of $\vt$ the character of motion is dramatically changed. For example, at
$\vt=20^\circ$ the particle is trapped by the magnetic field (see Fig.~\ref{xyz20}).
There is no sharp transition between these two cases. At some angles close
to $\vt=20^\circ$, the particle may make several turns around the star, with a quite
chaotic features of motion, and then go to infinity.

At large angles the trajectories are again changed quantitatively, and appear close to the field lines
(see Fig.~\ref{xyz30}). In all cases the initial segment of trajectory appears close to the field lines;
the acceleration on that segment is shown in Fig.~\ref{ta}. Therefore, for the chosen initial conditions
the curvature radius of the trajectory significantly differs from the curvature radius of the field line.
The initial segments of the trajectory are helical-like lines, but at $\xi\sim
10^{-3}$ the shift in the perpendicular direction appears of the order of
$10^{-3}$. Therefore within this uncertainty, the trajectories of particles
shown in Figs.~\ref{xyz10} -- \ref{xyz30} coincide with the smooth trajectories.

%\bibliography{cites}


\end{document}